# Crossover from charge order to strain glass in phase separated manganite thin films: Impact of thermal cycling and substrate induced strain

Vasudha Agarwal, P. K. Siwach, V. P. S. Awana, and H. K. Singh[#]
CSIR-National Physical Laboratory, Dr. K. S. Krishnan Marg, New Delhi-110012, India

Abstract

The magnetic and magnetotransport properties of single crystalline $La_{1-x-y}Pr_yCa_xMnO_3$ ($x\approx0.42$, $y\approx0.40$) thin films (~140 nm) deposited on (110) oriented $LaAlO_3$ and $SrTiO_3$ substrates exhibit a crossover from the high temperature antiferromagnetic-charge ordered insulator (AFM-COI) phase ($T>T_N$) to strain glass ($T<T_g$). At intermediate temperatures ($T_g \leq T \leq T_N$) dynamical liquid having prominent thermal-magneto-resistive hysteresis dominates in the cooling cycle, while in the warming cycle it is preceded by ferromagnetic metal (FMM) phase. Magnetic field required to drive AFM-COI to FMM phase transition are higher than that for the strain glass. The magneto-electric nature and temperature span of the distinct magnetic regimes are sensitive to the thermal cycling and substrate induced strain.

[#] Corresponding Author; Email: hks65@nplindia.org

The composition-temperature (x-T) phase diagram of doped rare earth manganites of the type $RE_{1-x}AE_xMnO_3$ (RE = Pr, Nd, La, etc. and AE = Ca, Sr, Ba, etc.) acquires higher degree of complexity at reduced $e_g$ electron bandwidth, which in turn is controlled by the average cationic radius $\langle r_{RE} \rangle$ of the RE/AE cations or the tolerance factor (t).[1-4] At lower values of $\langle r_{RE} \rangle$ the interaction between the various degrees of freedom, e.g., spin, charge, lattice, etc. is altered through the modification in the basic interactions, such as, the ferromagnetic double exchange (FM-DE), antiferromagnetic super exchange (AFM-SE), electron-lattice coupling through Jahn-Teller (JT) distortion, etc.[1-5] At reduced bandwidth the charge ordered insulator (COI) phase having AFM spin configuration acquires prominence over an appreciable range of x and hence the boundaries between the electronic phases, viz. paramagnetic insulator (PMI), ferromagnetic metal (FMM), ferromagnetic insulator (FMI), and AFM-COI become diffuse.[6-8] The ensuing competitive coexistence of phases leads to electronic phase separation; the most striking intrinsic feature of manganites.[1,2]

$La_{1-x-y}Pr_yCa_xMnO_3$ (x>0.3, y>0.4) has emerged as an ideal system for probing various aspects of phase separation and hence has been studied extensively with different combination of x and y in different forms.[9-17] Uehara et al.[9] have demonstrated the coexistence of submicron size FMM and AFM-COI clusters and the consequent percolative nature of the electrical transport. The AFM-COI phase is observable explicitly only at x>0.3, e.g., at y≈0.4 and x≈0.375 bulk polycrystals show the CO transition at $T_{CO}$ ≈230 K is followed by an AFM spin ordering around $T_N$ ≈ 180 K and finally the AFM-COI to FMM transition occurs at $T_C$ ≈ 80 K.[9-11] The free energy difference between FMM and AFM-COI phases is delicate, so much so, that their relative fraction changes rapidly with (i) time, (ii) temperature and (iii) little variation in x and y.[10,11] Just below the $T_C$, a dynamical liquid like phase appears and it persists over a fairly long temperature range till it gets transformed into a randomly frozen strain glass (SRG) like state at the glass transition temperature $T_g$.[11]

In magnetic and electrical transport measurements three distinct regions have been recognized in $La_{1-x-y}Pr_yCa_xMnO_3$.[9-17] The first appears at $T>T_C$ and has AFM-COI as the dominant phase. The second, referred to as a dynamical liquid like phase showing hysteretic first order insulator-metal transition (IMT) at $T_{IM}$ appears in the range $T_g<T<T_C/T_{IM}$. The third distinct region is the randomly frozen glassy state akin to SRG characterized by thermally reversible magnetization/resistivity and appears at $T<T_g$.[11] Since the nature of the electronic phase mixture in the three regimes is distinct, magnetotransport properties have contrasting temperature (T) and magnetic field (H) dependence. However, the T and H evolution of magnetotransport as the system crosses over from one regime to the other has not been adequately investigated in single crystalline thin films. Further, the nature of the dynamical magnetic liquid state during cooling and warming cycles has not been probed. In view of the delicate balance between the free energies of the coexisting phases, even a subtle strain originating from the substrate can appreciably alter the relative phase fractions and affect magnetotransport. The present work demonstrates that in single crystalline $La_{1-x-y}Pr_yCa_xMnO_3$ ($x\approx0.42$, $y\approx0.40$) thin films the FMM, AFM-COI, the dynamical liquid like phase and SRG are sensitive to thermal cycling and substrate driven strain. It has also been demonstrated that the magnetic liquid phase observed at $T_g<T<T_C$ in the cooling cycle transforms first into a crystalline FMM phase and then to dynamical liquid like phase during the warming cycle. The temperature dependence of AFM-COI/SRG melting field clearly shows a cross-over minimum which is broadened by the compressive strain.

$La_{1-x-y}Pr_yCa_xMnO_3$ ($x\approx0.42$, $y\approx0.40$) thin films (~140 nm) were grown by RF magnetron sputtering of a stoichiometric (2 inch) target in 200 mtorr of Ar + $O_2$ (80 + 20) mixture on (110) oriented single crystal $SrTiO_3$ (STO) and $LaAlO_3$ (LAO) substrates maintained at ~800 °C. All films were annealed at $\approx$900 °C for 10 hr in flowing oxygen. The high resolution X-ray diffraction (HRXRD; $2\theta$-$\omega$ scan, $\phi$-scan and $\omega$-$2\theta$ scan) confirmed the epitaxial nature of the films and also showed that the film on LAO (referred hereafter as L-LPCMO) is under compressive strain, while that on STO (S-LPCMO) is under tensile strain. Temperature dependent magnetization (M-T) was measured by SQUID magnetometer (MPMS-Quantum Design) employing zero field cooled (ZFC), field cooled cool (FCC) and field cool warming (FCW) protocols. The magnetotransport was measured by a physical property measurement system (PPMS-Quantum Design).

The M-T data plotted in Fig. 1(a,b) reveals multiple transitions in both films. The transition temperatures were determined from the ZFC $\frac{dM}{dT}$ curves shown in the respective insets. The change in the slope of the M-T curves at $T_{CO}^{ON}\approx230$ K and $\approx275$ K in L-LPCMO and S-LPCMO, respectively could be attributed to the appearance of CO correlations. The inverse peak in $\frac{dM}{dT}$ data yields $T_{CO}\approx185$ K and $\approx235$ K in L-LPCMO and S-LPCMO, respectively. On lowering T the magnetization of L-LPCMO continues to rise. This suggests quenching of the AFM transition due to compressive strain. However, the $\frac{dM}{dT}$ data suggests AFM transition at $T_N\approx157$ K. In contrast, S-LPCMO shows a decrease in the magnetization as T lowered below $T_{CO}$ and undergoes AFM transition at $T_N\approx184$ K. The rise in magnetization at $T_{FM}^{ON} \approx$ 129 K (L-LPCMO, $T_{FM}^{ON}$ is the temperature from where FM cluster starts growing) and ≈121 K (S-LPCMO) could be taken as the appearance of FM phase in the AFM-COI matrix. ZFC, FCC and FCW data of L-LPCMO yielded three distinct $T_C$ values for the three measurement protocols; $T_C^{ZFC} \approx$ 100 K, $T_C^{FCC} \approx$ 65 K and $T_C^{FCW} \approx$ 102 K. S-LPCMO shows $T_C^{ZFC} \approx$ 94 K, $T_C^{FCC} \approx$ 48 K and $T_C^{FCW} \approx$ 96 K. At $T<T_C$ the ZFC M-T curve shows a sharp decline at $T_P \approx$ 52 K, while a peak occurs in the FCW curve simultaneously in both the films. This could be regarded as the onset of the freezing of the dynamical liquid like magnetic phase. The slope change at $T_g \approx$ 28 K in both the films could be regarded as the transformation of the dynamical liquid to SRG. The discussion presented above clearly shows that AFM-COI phase is more sensitive to substrate induced strain than the FM transition.

At $T \leq T_{CO}$, ZFC and FCW M-T curves of both L-LPCMO and S-LPCMO show prominent divergence, generally attributed to a metamagnetic cluster glass.[18-20] The strong FCC-FCW hysteresis ($T_g < T < T_{FM}^{ON}$) and the associated large difference between the $T_C^{FCC}$ and $T_C^{ZFC}/T_C^{FCW}$ is due to the supercooling of the magnetic liquid.[16,17] The force driving supercooling is magnetic frustration resulting from the competitive coexistence of FM and AFM-COI phases, which hinders the nucleation of FM spin order at the equilibrium $T_C$. At $T<T_g \approx$ 28 K, the FCC-FCW curves exhibit thermal reversibility and weak temperature dependence. This temperature regime is inhabited by randomly frozen SRG. The higher magnetic moment and $T_C$ values and suppression of AFM-COI phase shows that compressively strained L-LPCMO has higher FM fraction than S-LPCMO.

None of these films show any signature of the CO transition in the cooling cycle temperature dependent resistivity (ρ-T) data (Fig. 2). L-LPCMO shows a sharp IMT at $T_{IM}^C$ ≈64 K and at T<25 K the resistivity appears to saturate. In the warming cycle ρ-T remains reversible till about T≈25 K and IMT appears at $T_{IM}^W$ ≈116 K. In cooling cycle ρ-T of S-LPCMO becomes immeasurable at T<61 K but around 30 K it shows huge drop by more than five orders of magnitude. During warming cycle, ρ-T shows a reversible behaviour till about T~20 K and then shows a minimum at $T_m \approx$ 34 K, which is followed by an IMT at $T_{IM}^W$ ≈96 K. In warming cycle, CO transition appears at $T_{CO} \approx$ 232 K and ≈238 K in L-LPCMO and S-LPCMO, respectively. The sharp-hysteretic IMT observed in both films is also representative of the first order phase transition.[9,16,17] The sharp drop in ρ-T during the cooling cycle could be regarded as the representative of the dynamical liquid behaviour and the saturation of ρ-T at $T<T_g$ represents the crossover to the SRG regime. During warming cycle the initial slow increase and the subsequent sharp rise in the ρ-T curve could be taken as the signature of a crystalline FMM solid and magnetically frustrated dynamical liquid, respectively. The thermally reversible behaviour at $T<T_g$ is a manifestation of the SRG, while the minimum at $T_m$ could be regarded as the thermal devitrification of SRG during warming. Application of magnetic field enhances $T_{IM}^C/T_{IM}^W$, reduces the ρ–T hysteresis, and dilutes the first order nature of the IMT (inset of Fig. 2). The two distinct transitions at $T_{IM}^C$ and $T_{IM}^W$ in the cooling and warming cycle are attributed to the supercooling and superheating of the magnetic liquid consisting of FMM and AFM-COI phases.[16,17]

To evaluate the impact of magnetic field on electrical transport in the different magnetic regimes discussed above we measured the magnetic field dependent resistivity (ρ-H) at several temperatures (5 K ≤ T ≤ 200 K) and the representative ρ-H data of the two films are presented in Fig. 3. At T=5 K, virgin cycle ρ-H curves of both the films shows sharp decrease when H exceeds a certain value ($H_m$), the original ρ (H=0) is never recovered in subsequent cycles and ρ (H) decrease continuously resulting in non-hysteretic behaviour. As H is increased from $H_m$ to 50 kOe, ρ (H) decreases by ~50% in L-LPCMO and by almost three orders of magnitude in S-LPCMO. Here $H_m$ could be regarded as the melting field to achieve FM order and its value at T=5 K is found to be ≈25 kOe and ≈27 kOe for L-LPCMO and S-LPCMO, respectively. The observed behavior at T=5 K is due to the H induced melting of the magnetically disordered SRG to form crystalline FMM. At T=25 K L-LPCMO shows very small decrease in ρ (H) during the virgin cycle and strong hysteresis is observed in subsequent cycles. The decrease in ρ (H) and the associated hysteresis is more pronounced in S-LPCMO. The value of $H_m$ is found to be ≈18 kOe and ≈20 kOe for the L-LPCMO and S-LPCMO films, respectively. The ρ-H hysteresis confirms FMM and AFM-COI coexistence and is a consequence of different dynamics of AFM (FM) to FM (AFM) during H increasing (decreasing) cycle.

As discussed earlier the temperature regime immediately above $T_g$ ($T_g < T < 75$ K and $T_g < T < 90$ K for S-LPCMO and L-LPCMO, respectively) consists dominantly fraction of FMM with small fraction of AFM-COI. Due to this the value of $H_m$ as well as the drop in ρ (H) is very small. As temperature is raised further up the fraction of AFM-COI phase increases rapidly. This gives rise to a magnetically frustrated dynamical liquid like phase, which shows a multi-order decrease in resistivity even at moderate magnetic fields. This is clearly demonstrated by the ρ-H curves of L-LPCMO and S-LPCMO measured at 125 K and 100 K, respectively. In this temperature regime values of $H_m$ are moderate, being ≈ 17 kOe and 16 kOe for the L-LPCMO and S-LPCMO, respectively. At $T>T_{FM}^{ON}$, which is dominated by AFM-COI phase $\rho$ (H) after initial slow decrease drops sharply at $H>H_m$. This is due to the H induced melting of the AFM-COI phase. This is clearly demonstrated by $\rho$-H curve of the L-LPCMO measured at 150 K (Fig. 3a) and the S-LPCMO measured at 125 K and 150 K (Fig. 3b). In case of L-LPCMO the melting field is found to be $H_m \approx 36$ kOe and the same for S-LPCMO is found to be $\approx 29$ kOe and $\approx 41$ kOe at 125 K and 150 K, respectively.

The variation of $H_m$ with T (Fig. 4) clearly shows the existence of four temperature regimes inhabited by different magnetic phases. The higher temperature region ($T_C \leq T \leq T_{CO}$) is dominantly AFM-COI with the fraction of FMM increasing as T approaches $T_C$. This region is characterized by the highest $H_m$. As the temperature is lowered the AFM-COI (FMM) fraction decreases (increases) and hence the observed gradual decrease in $H_m$. Just below $T_C/T_{IM}$ the coexisting phases are delicately balanced and magnetic frustration is predominant. This favors the dynamical liquid like phase. Moderate magnetic fields are sufficient to drive the disordered liquid phase into a FMM one. The valley region with the lowest values of $H_m$ corresponds to the crystalline FMM phase with small fraction of embedded AFM-COI. As the temperature is lowered further, the magnetic liquid freezes gradually (freezing begin at $T_P \approx 52$ K) to SRG with $T_g \approx 28$ K. The disordered nature of SRG requires higher $H_m$ to transform the glass into a crystalline FMM phase. We have also analysed the temperature dependence of change of $\rho$ as H is increased from $H_m$ to H=50 kOe (inset of Fig. 4). This also shows four distinct temperature regimes. The divergence between $\rho$ ($H_m$) and $\rho$ (50 kOe) below $T_g$ is an unambiguous signature of the glassy phase and the same above $T_C/T_{IM}$ shows the dominance of the AFM/COI phase. The small difference in $\rho$ ($H_m$) and $\rho$ (50 kOe) at $T_g < T < 75$ K/90 K is due the dominance of the FMM phase and the strong divergence just below $T_C$ is a manifestation of the dynamical liquid like phase. The impact of strain is reflected by the fact that throughout the temperature range, $H_m$ is smaller (larger) for the compressively (tensile) strained L-LPCMO (S-LPCMO). The broadening of the minimum of the T-$H_m$ in L-LPCMO could be regarded as a signature of the higher FMM fraction, which as pointed out earlier is a consequence of compressive strain in this film. The fact that the $\rho$ (H) divergence is more pronounced in S-LPCMO clearly brings out the effect of substrate induced strain.

M-T, $\rho$-T and $\rho$-H data presented above confirms the strongly phase separated nature and establish the existence of multiple magnetic regimes in La$_{1-x-y}$Pr$_y$Ca$_x$MnO$_3$ (x$\approx$0.42, y$\approx$0.40) thin films. The degree of phase separation and the temperature span of magnetic regimes are sensitive to thermal cycling and substrate driven strain. In cooling cycle the high temperature AFM-COI phase is separated from the low temperature SRG by the dynamical magnetic liquid phase. In warming cycle the SRG undergoes devitrification and the dynamic liquid is preceded by a crystalline FMM phase. High magnetic fields ($H_m$) are required to melt the SRG and the AFM-COI dominated regime, while the liquid like phase is the most sensitive to the external perturbations.

**Acknowledgements**

Authors are grateful to Prof. R. C. Budhani for persistent encouragement. Financial support from CSIR is thankfully acknowledged. VA is grateful to CSIR for the award of a senior research fellowship. Dr. Anurag Gupta is thankfully acknowledged for magnetic measurements (MPMS-DST facility).

**Figure Captions**

Figure 1  Temperature dependent ZFC, FCC and FCW magnetization of (a) L-LPCMO and (b) S-LPCMO measured at H=100 Oe. The respective insets show $\frac{dM}{dT} - T$ plot of ZFC data.

Figure 2  The ρ-T plot of LPCMO films on LAO and STO substrates measured during cooling (C) and warming (W) cycles. The inset shows ρ-T measured at H=50 kOe magnetic field.

Figure 3  The representative ρ-H plots of (a) L-LPCMO and (b) S-LPCMO measured at different temperatures.

Figure 4  The observed temperature dependence of $H_M$ in S-LPCMO and L-LPCMO. The inset shows the variation of ρ of the two films measured at $H_m$ and H=50 kOe.

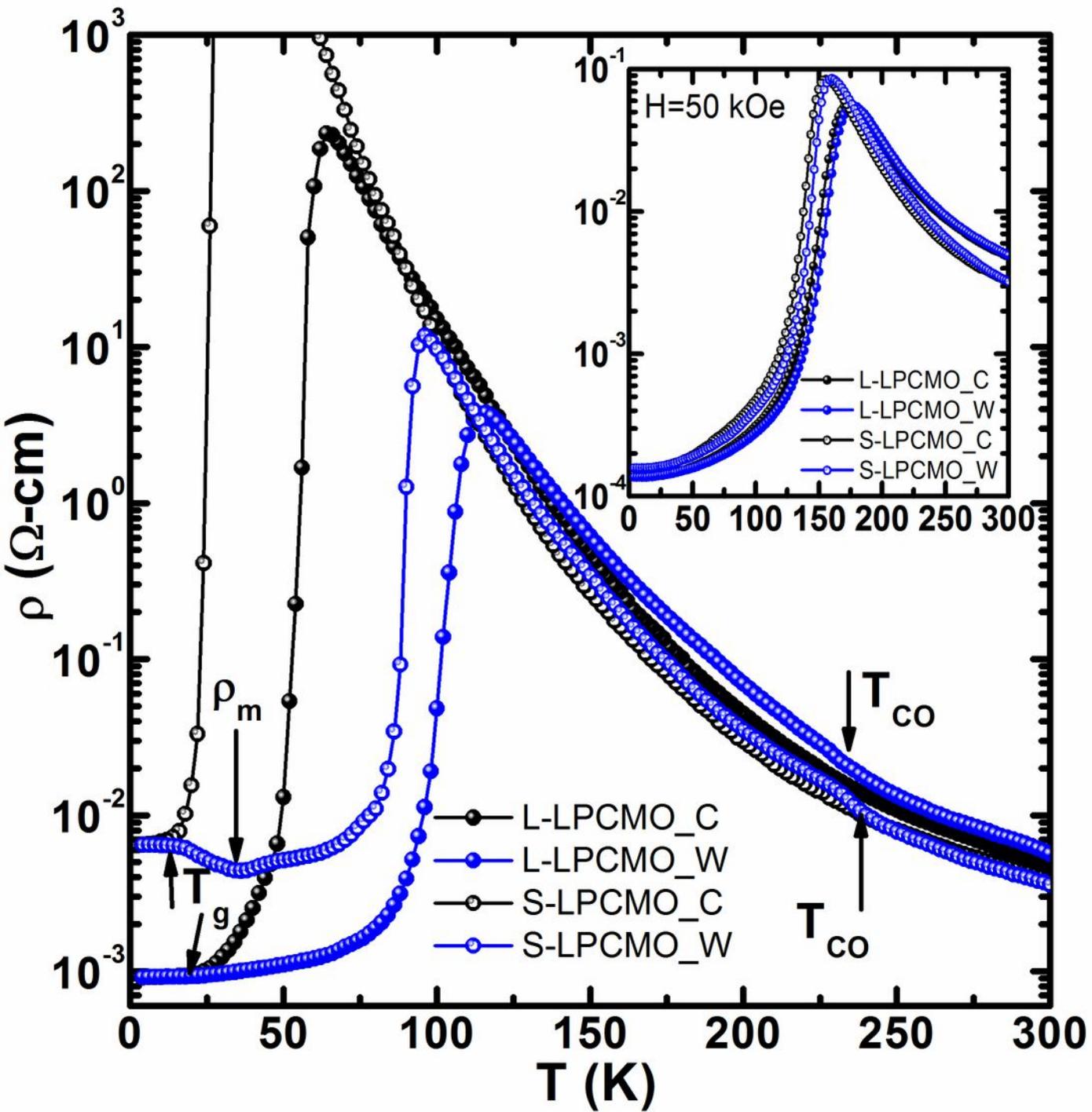

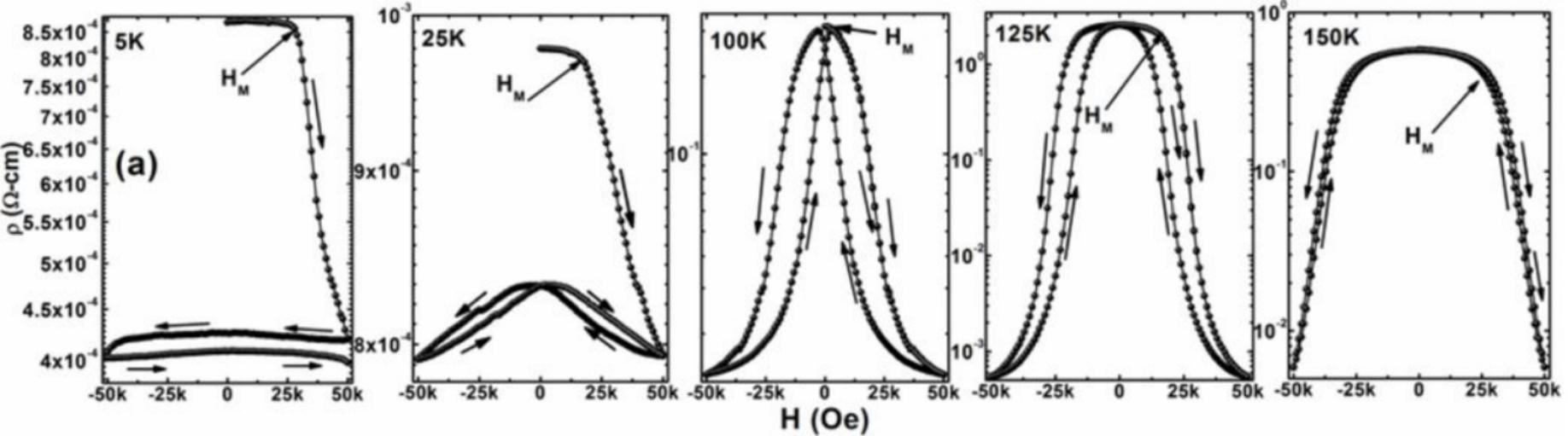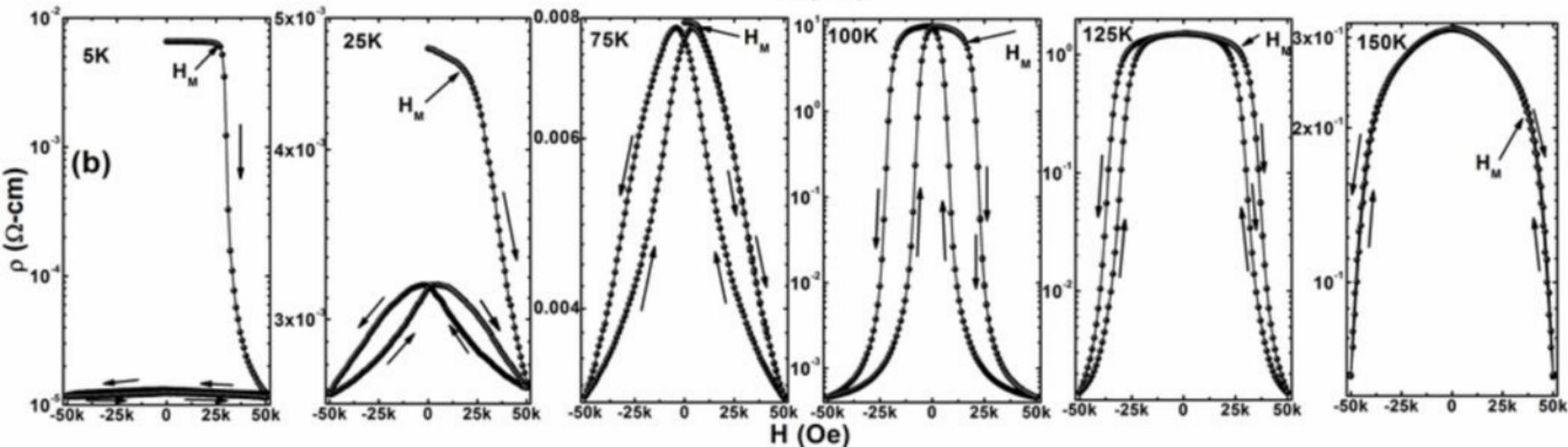

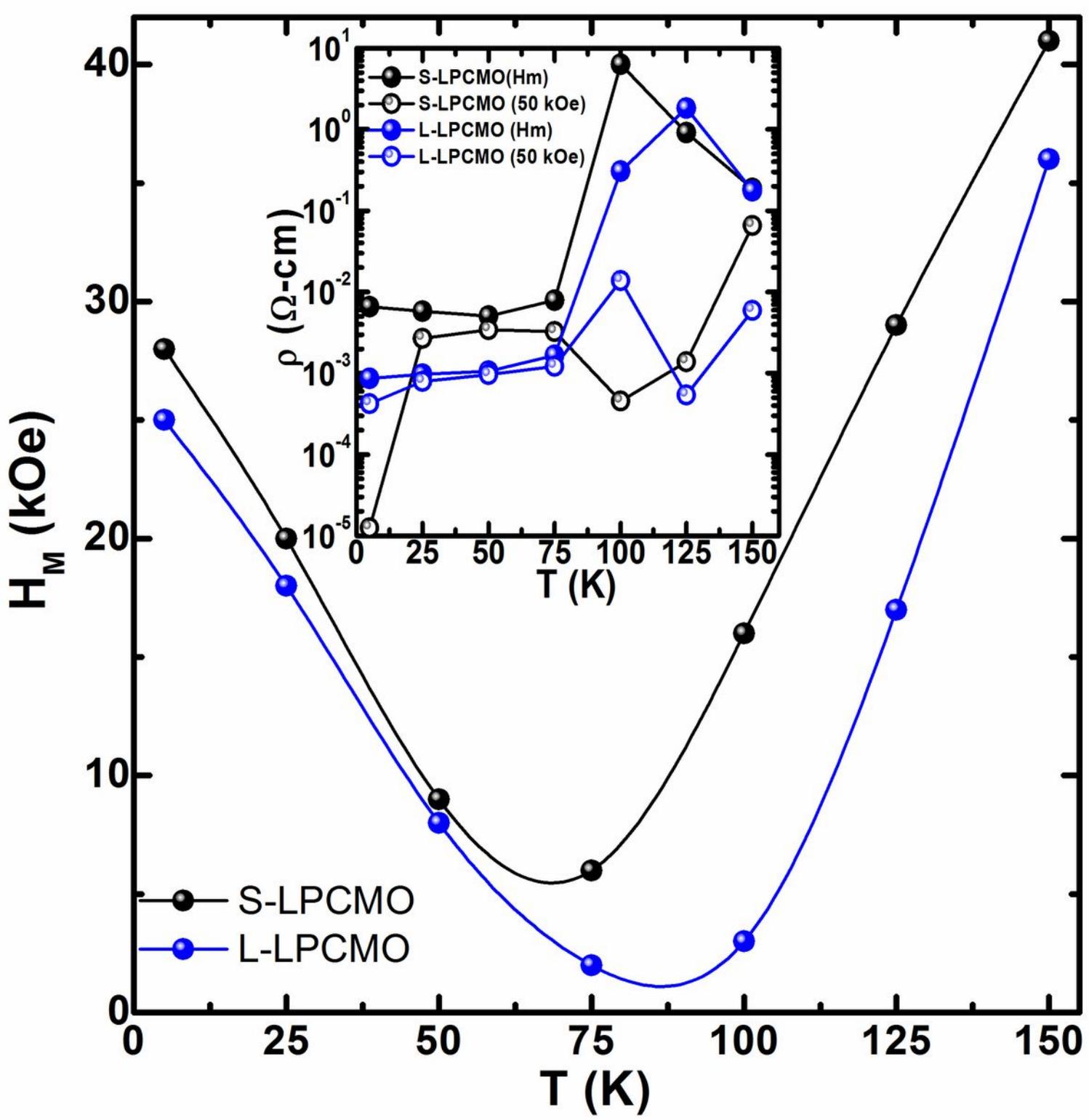